\documentclass[10pt,conference]{IEEEtran}

\usepackage{acronym}
\usepackage[]{float}
\usepackage{listings,xfp}
\usepackage{soul}
\usepackage{subcaption}
\usepackage{tikz}
\usetikzlibrary{decorations,chains,shapes,positioning,automata,arrows,arrows,calc,arrows.meta}
\usepackage{xspace}
\usepackage{amssymb}
\usepackage{amsmath}
\usepackage{amsthm}
\usepackage{relsize}
\usepackage{anyfontsize}
\usepackage{epstopdf}
\usepackage{tabularx}
\usepackage{graphics}
\usepackage{enumitem}
\usepackage[english]{babel}
\newtheorem{theorem}{Theorem}

\usepackage[T1]{fontenc}
\usepackage{newtxtext,newtxmath} 

\usepackage{soul}
\usepackage{cite}
\usepackage{url}
\usepackage{listings}
\usepackage{xcolor}
\lstdefinestyle{codeC}{
	language=C,
	numbers=right,
	numberstyle=\footnotesize,
	numbersep=6pt,
	frame=single,
	breaklines=true,
	columns=fullflexible,
	basicstyle=\linespread{1.0}\ttfamily \footnotesize,
	xleftmargin=0.5em,
	framexleftmargin=-0.25em,
	xrightmargin=4pt,
	tabsize=1,
	showstringspaces=false,
	captionpos=b,
	morekeywords={BitHash1, BitHash2}
}

\lstdefinelanguage{CUDA}{
	language=C++,
	morekeywords={__device__,__host__,__forceinline__,__global__,__restrict__,warp,threadIdx,blockIdx},
	sensitive=true
}
\usepackage[ruled,vlined,linesnumbered]{algorithm2e}

\makeatletter
{\small 
  \xdef\f@size@small{\f@size}
  \xdef\f@baselineskip@small{\f@baselineskip}
  \normalsize 
  \xdef\f@size@normalsize{\f@size}
  \xdef\f@baselineskip@normalsize{\f@baselineskip}
}
\newcommand{\CodeSize}{%
  \fontsize
  {\fpeval{(\f@size@normalsize)*0.75}}
  {\fpeval{(\f@baselineskip@normalsize)*0.75}}%
  \selectfont
}
\makeatother

\makeatletter
\renewcommand\subsubsection{\@startsection{subsubsection}{3}{\z@}%
	{0pt}
	{0.5ex \@plus .2ex}
	{\normalfont\normalsize\bfseries}} 
\makeatother

\SetKw{Continue}{continue}
\SetKw{Broadcast}{broadcast}
\SetKwBlock{KwFn}{function}{end}
\newcommand{\PlacedWithoutEvict}{\textsc{PlacedWithoutEvict}}

\newcommand{\var}[1]{\texttt{#1\xspace}}
\newcommand{\fig}[1]{\mbox{Figure~\ref{fig:#1}}\xspace}

\newcommand{\hivehash}{Hive hash table\xspace}

\acrodef{afa}[AFA]{Atomic Fetch Add}

\acrodef{axchg}[XCHG]{Atomic Exchange}

\acrodef{rmw}[RMW]{Read-Modify-Write}

\acrodef{rss}[RSS]{Residual Sum of Squares}

\acrodef{cas}[CAS]{Compare and Swap}

\acrodef{cu}[CU]{Compute Unit}

\acrodef{fifo}[FIFO]{First-In-First-Out}

\acrodef{GPU}[GPU]{Graphics Processing Unit}

\acrodef{SIMT}[SIMT]{Single Instruction, Multiple Threads}
\newcommand{\simt}{\ac{SIMT}\xspace}

\acrodef{ISA}[ISA]{Instruction Set Architecture}

\acrodef{OS}[OS]{Operating System}

\newcommand{\BigO}[1]{$\mathcal{O}\left(#1\right)$}

\definecolor{ReadOnlyColor}{RGB}{192,192,192}

\begin{document}
  \sloppy
  
  \title{Hive Hash Table: A Warp-Cooperative, Dynamically Resizable Hash Table for GPUs}
	\author{
		\IEEEauthorblockN{Md Sabbir Hossain Polak\IEEEauthorrefmark{1}, David Troendle\IEEEauthorrefmark{2}, Byunghyun Jang\IEEEauthorrefmark{2}}
		\IEEEauthorblockA{\IEEEauthorrefmark{1}\IEEEauthorrefmark{2}Department of Computer and Information Science\\ University of Mississippi\\Oxford, MS, USA\\
		\{mhpolak@go, david@cs, bjang@cs\}.olemiss.edu}
	}

	  \maketitle
  
  	\begin{abstract}
  	Hash tables are essential building blocks in data-intensive applications, yet existing GPU implementations often struggle with concurrent updates, high load factors, and irregular memory access patterns. We present \textbf{\hivehash}, a high-performance, warp-cooperative and dynamically resizable GPU hash table that adapts to varying workloads without global rehashing.

  	\hivehash makes three key contributions. First, a cache-aligned packed bucket layout stores key–value pairs as 64-bit words, enabling coalesced memory access and atomic updates via single-CAS operations. Second, warp-synchronous concurrency protocols—Warp-Aggregated-Bitmask-Claim (WABC) and Warp-Cooperative Match-and-Elect (WCME)—reduce contention to one atomic operation per warp while ensuring lock-free progress. Third, a load-factor-aware dynamic resizing strategy expands or contracts capacity in warp-parallel K-bucket batches using linear hashing, maintaining balanced occupancy.

  	To handle insertions under heavy contention, \hivehash employs a four-step strategy: replace, claim-and-commit, bounded cuckoo eviction, and overflow-stash fallback. This design provides lock-free fast paths and bounded recovery cost under contention determined by a fixed eviction depth, while eliminating ABA hazards during concurrent updates.

  	Experimental evaluation on an NVIDIA RTX 4090 shows \hivehash sustains load factors up to 95\% while delivering 1.5–2× higher throughput than state-of-the-art GPU hash tables (SlabHash, DyCuckoo, WarpCore) under mixed insert–delete–lookup workloads. On balanced workload, \hivehash reaches 3.5 billion updates/s and nearly 4 billion lookups/s, demonstrating scalability and efficiency for GPU-accelerated data processing.
  	\end{abstract}


    \section{Introduction} \label{sec:Introduction}
    Modern GPUs have evolved far beyond their original graphics acceleration role to become powerful general-purpose computing platforms. They now deliver supercomputer-level performance across diverse domains including cryptocurrency mining \cite{dev2014bitcoin, iyer2018gpu}, network analysis \cite{shi2011fast, barrionuevo2015solving}, and database technologies \cite{bress2014gpu, rui2020efficient}, among others. These applications increasingly demand high-throughput, concurrent data structures \cite{moir2018concurrent, polak2024agile} that can exploit GPU architectures effectively.

    A common challenge in GPU computing is managing sparsely populated, irregular data domains. For example, particle tracking in computational fluid dynamics requires monitoring active cells in a large 3D grid where most cells remain empty. Dense array storage allocates memory for all grid cells -including empty ones - would be highly inefficient; instead, dynamic key-value structures are needed to represent only occupied regions. Prior work has explored spatial data structures such as Bounding Volume Hierarchies (BVHs) \cite{figueiredo2024bsh} and R-trees \cite{papadopoulos1997performance} for real-time queries, but these tree-based approaches incur \BigO{\log N} lookup costs that limit performance in latency-sensitive applications.
  
%
  
  In contrast, hash tables offer average-case $\mathcal{O}(1)$ query performance, making them a natural fit for real-time GPU applications. However, traditional schemes such as linear and quadratic probing suffer from clustering and irregular memory access, leading to poor memory coalescing and degraded throughput on GPUs. Cuckoo hashing mitigates clustering but introduces memory thrashing and may require global rehashing when eviction cycles occur. Existing GPU implementations (See \S\ref{sec:Related Works} Related Works.) further remain static and inflexible—either over-allocating memory for worst-case load or under-allocating and triggering costly global rehashing, which involves host-device transfers. As table occupancy (the fraction of filled slots) and update concurrency (simultaneous atomic inserts and deletes) increase, many GPU hash tables experience sharp throughput degradation or the design resorts to multiple subtables, which complicates lookups and raises synchronization overheads. These limitations motivate the need for a dynamic GPU hash table that can adapt to varying workloads, preserve coalesced access, and avoid global rehashing.

  In this paper, we propose \hivehash, a high-performance, dynamically resizable GPU hash table that supports fully concurrent mixtures of insertion, lookup, and deletion operations. Here, efficiency refers to both query throughput and memory utilization—that is, how effectively the implementation uses allocated GPU memory while sustaining high operation rates. Although dynamic resizing is often assumed to degrade query and build performance, \hivehash demonstrates that it can maintain high throughput for updates and lookups while preserving compact memory usage. As its probing scheme, we adopt cuckoo hashing for its $\mathcal{O}(1)$ worst-case lookup guarantee, enhanced with warp-cooperative concurrency control and bounded eviction.

  The followings summarize the key contributions of \hivehash:
  
  \begin{enumerate}
  	\item \textbf{Cache-Aligned Packed Buckets: } Each bucket, aligned to a cache line, stores key–value pairs as 64-bit packed words. Modern CUDA architectures natively support 64-bit atomic operations on global memory, allowing each packed entry to be updated or removed atomically with a single compare-and-swap (CAS). This compact array-of-structures (AoS) layout maximizes packing density and ensures fully coalesced memory access across warps.

  	\item \textbf{Warp-Cooperative Concurrency Protocols: } Two lightweight warp-synchronous primitives—Warp-Aggregated-Bitmask-Claim (WABC) for cooperatively slot allocation and Warp-Cooperative Match-and-Elect (WCME) for lookup, replacement, and deletion—aggregate contention into a single atomic action per warp, eliminating redundant operations and ensuring lock-free progress.

  	\item \textbf{Load-Aware Dynamic Resizing: } \hivehash dynamically grows or contracts the table in K-bucket batches using a warp-parallel variant of Linear Hashing. Each warp independently splits or merges partner buckets based on load factor thresholds, allowing incremental expansion without global rehashing. This strategy maintains high occupancy while keeping resizing cost amortized and predictable.
  \end{enumerate}

  \section{Related Works} \label{sec:Related Works}
  GPU hash tables have evolved through several generations, each addressing specific performance and scalability challenges. Early static designs achieved high throughput but suffered from inflexibility, while recent dynamic approaches have struggled to maintain performance under concurrent updates (as shown in \fig{imbalance_workload_perf}).

  \textbf{Static GPU Hash Tables.} Alcantara et al. \cite{alcantara2012building} pioneered cuckoo hashing on GPUs, demonstrating high load factors but requiring preallocated memory. Subsequent work optimized specific aspects: StadiumHash \cite{khorasani2015stadium} and MegaKV \cite{zhang2015mega} improved warp-cooperative probing and memory coalescing, while cuDF \cite{cuDF} added multi-value support through linear probing. However, linear probing suffers from clustering under skewed distributions, and HashGraph \cite{green2021hashgraph} adopted sparse matrix layouts that incur significant memory overhead. These static approaches prioritize speed over memory efficiency, making them unsuitable for dynamic workloads or memory-constrained environments where multiple data structures must coexist.

  \textbf{Dynamic GPU Hash Tables.} Supporting concurrent insertions and deletions on GPUs introduces fundamental challenges in memory management and consistency. SlabHash \cite{ashkiani2018dynamic} pioneered dynamic GPU hashing through linked ``slabs" that enable on-demand growth. However, this design incurs pointer-chasing overhead and requires symbolic deletion markers that cause memory bloat. Under high contention, slab traversal exhibits $\Omega(\log \log m)$ complexity \cite{ashkiani2018dynamic}, degrading performance for mixed insert/delete workloads. The fundamental challenge is maintaining both constant-time operations and memory efficiency under concurrent modifications.

  \textbf{Incremental Resizing Approaches.} Recent work has explored alternatives to full-table rehashing. DyCuckoo \cite{li2021dycuckoo} introduces multi-subtable resizing that expands or contracts only affected partitions, maintaining stable load factors but adding lookup overhead since queries must probe multiple independent subtables. Iceberg hashing \cite{bender2023iceberg} demonstrates the theoretical difficulty of combining high space efficiency, update performance, and constant-time guarantees. On GPUs, these challenges are amplified by memory bandwidth constraints and the cost of large-scale data movement.

  \textbf{Positioning of HiveHash.} Our work addresses these limitations through three key innovations described in Section \S \ref{sec:Introduction}. Unlike static designs, \hivehash supports dynamic resizing without sacrificing constant-time lookup guarantees. Unlike slab-based approaches, it avoids allocator overhead through direct bucket management and immediate slot reuse. Unlike multi-subtable designs, HiveHash maintains a single logical address space while incrementally adjusting capacity through warp-parallel linear hashing. This combination enables both high performance and memory efficiency under mixed concurrent workloads.
  
  \section{\hivehash} \label{sec:HiveHash}
  \hivehash is an open-addressing hash table that stores entries as 64-bit key-value (KV) words (32-bit key, 32-bit value) and executes concurrent operations in a single monolithic GPU kernel.
  In this section, we introduce the packed AoS bucket layout (\S\ref{subsec:Bucket Layout}), describe the design of the \hivehash structure (\S\ref{subsec:HashtableDS}), outline the hash functions policy (\S\ref{subsec:hashfunctions}), and discuss additional design considerations.

  \subsection{Memory Layout and Bucket Design} \label{subsec:Bucket Layout}
\begin{figure}[!t]
  	\centering
  	\begin{subfigure}[t]{0.5\linewidth}
  		\centering
  		\includegraphics[width=0.85\linewidth]{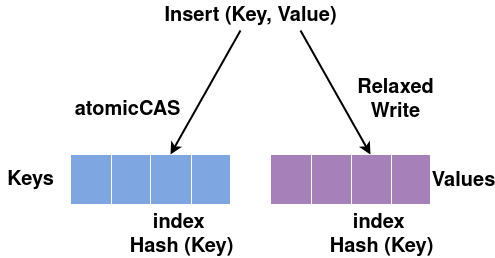}
  		\caption{SoA Memory Layout}
  		\label{fig:soa-memory-layout}
  	\end{subfigure}\hfill
  	\begin{subfigure}[t]{0.5\linewidth}
  		\centering
  		\includegraphics[width=0.95\linewidth]{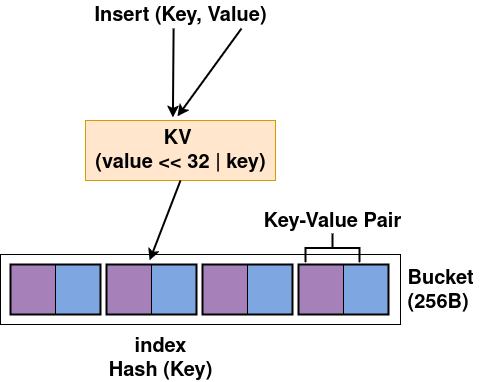}
  		\caption{AoS Memory layout with Packed 64-bit KV Pair}
  		\label{fig:aos-bucket-memory-layout}
  	\end{subfigure}
  	\caption{Bucket Memory Layout}
  	\label{fig:bucket-memory-layout}
  \end{figure}

  In \hivehash, a bucket is the local collision window: all keys mapped by the hash functions to the same index compete for the same bucket slots. On \simt hardware, multiple threads can target the same bucket concurrently; thus the layout must maximize coalescing and minimize contention among concurrent updates. A classical structure-of-arrays (SoA) maintains distinct arrays for keys and values, which avoids interleaving but incurs \emph{two-phase updates}: one atomic CAS to claim the key slot and a subsequent relaxed write to publish the value, creating extra traffic to global memory (\fig{soa-memory-layout}).
  
	Instead, \hivehash adopts a \emph{Packed Array-of-Structure (AoS)} organization as shown in \fig{aos-bucket-memory-layout}: each bucket is a single linear array of 64-bit KV pairs as encoded as follows.
	
	\[
	\texttt{pair} = (\texttt{(value)} << 32) \;|\; \texttt{key}
	\]
	Key and value can be extracted with a simple bit-level operation as follows.
	\[
	\texttt{key} = \texttt{pair} \ \&\ 0xFFFFFFFF\texttt{u},
	\]
	\[ 
	\texttt{value} = \texttt{pair} >> 32 
	\]
	This compact representation preserves AoS's single index space while collapsing each entry into one word. It enables a \emph{single 64-bit CAS} to publish or remove both fields atomically for modern GPUs, eliminating the need of \emph{CAS+store}. It removes the key/value inconsistency and simplifies correctness in the presence of concurrency.

 	Buckets are sized and aligned for warp-level coalescing. Modern NVIDIA GPUs utilize 128-byte cache lines for both L1 and L2 cache\cite{NVIDIA_Memory_Coalescing}. We provision 32 slots per bucket so a warp of 32 threads can probe a bucket in parallel with one lane per slot. With 64-bit entries, a full bucket occupies 256 bytes and is aligned so any probe touches at most two cache lines. For configurations using 32-bit entries (e.g., key-only workloads or 16-bit key and 16-bit value), each bucket fits within a single cache line. In both configurations, the packed AoS layout yields coalesced accesses for probes and avoids the second memory transaction required in classical SoA layouts.
 	
 	\subsection{Hash Table Data Structure} \label{subsec:HashtableDS}
\begin{figure}[!ht]
	\centering
	\scalebox{0.35}
	{\includegraphics{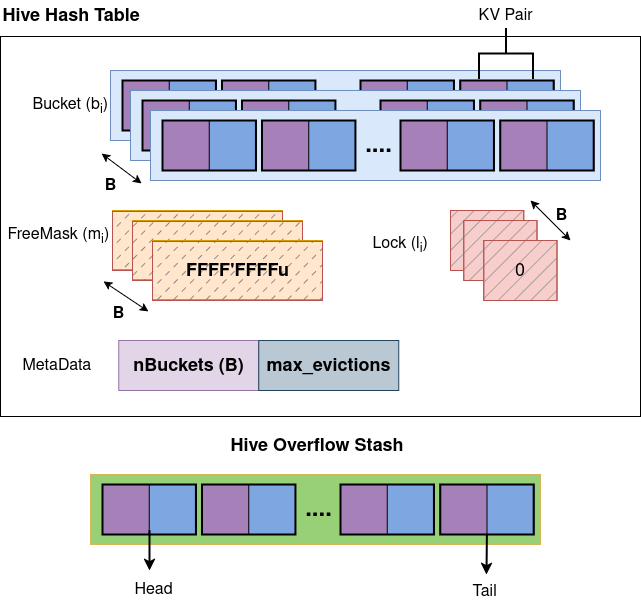}}
	\caption{\hivehash data structure.}
	\label{fig:HashTableDataStructure}
 	\end{figure}

 	\fig{HashTableDataStructure} illustrates the core organization of \hivehash. The design decouples the bucket array from auxiliary metadata arrays (\texttt{freeMask}, \texttt{lock}) and global control fields to maximize memory coalescing, simplify warp-synchronous operations, and enable predictable access patterns. 
 	
%
 	
	\begin{itemize}
	\item \textbf{Buckets (\(b\)): } The slot array of buckets stores the packed 64-bit key-value words. Each bucket holds a fixed number of slots ($S=32$ by default) and is aligned for warp-level coalesced access to all lanes of a warp.

	\item \textbf{Free Mask (\(m\)): } Each bucket is associated with a 32-bit occupancy bitmap, where bit $i$ indicates whether slot $i$ is available (1) or occupied (0). A warp can check availability with a single load rather than scanning all KV pairs. Claiming a slot requires just one read-modify-write (RMW) on the mask, followed by accessing a 64-bit packed KV stored in the bucket slot. This design provides constant-time, contention-resilient slot allocation.

	\item \textbf{Lock (\(l\)): } A boolean reserved per bucket that locks bucket access. It is used only in rare eviction cases where both candidate buckets are full and an entry must be displaced. Regular operations—insert, lookup, and delete—bypass the lock entirely, maintaining a lock-free fast path even at high load factors (used in $< 0.85\%$ of cases, as shown in \S\ref{subsec:insertion_breakdown}).

	\item \textbf{Global Metadata: } A small set of scalar values tracks global state: $nBuckets$ denotes the current number of buckets and a parameter $max\_evictions$ bounds the length of cuckoo displacement chains. When this bound is exceeded, the operation falls back to the overflow stash.
	
	\item \textbf{Overflow Stash: } Insertions that fail after exhausting both candidate buckets and the eviction bound are redirected to a small auxiliary buffer called the \textit{Overflow Stash}. Implemented as a lock-free queue in global memory, it temporarily stores overflow items, which are later reinserted after the next resizing phase to preserve correctness and progress.
	\end{itemize}
  
\subsection{Hash Functions} \label{subsec:hashfunctions}
\hivehash requires efficient mapping of keys to buckets. Formally, let $K$ denote the key space, $B=\{0,1,\dots,|B|-1\}$ the bucket indices, and $h:K \rightarrow B$ a hash function. At a minimum, $h$ must satisfy the following properties~\cite{korwar2010universal}:

\begin{enumerate}
	\item \textbf{Well-defined mapping.} $\forall k \in K$, $h(k)$ deterministically produces a unique $b \in B$; i.e., repeated evaluations of $h(k)$ are independent of past history.
	\item \textbf{Uniform distribution.} Each bucket is selected with probability $1/|B|$, ensuring balanced occupancy in expectation.
	\item \textbf{Key-order independence.} Uniformity must hold regardless of input distribution or ordering (e.g., odd keys should not bias toward odd buckets).
\end{enumerate}

\begin{lstlisting}[
	style=codeC,
	caption={BitHash1 and BitHash2 Hash Functions.}]
	uint32_t BitHash1(uint32_t key, uint32_t nBuckets) 
	{
		key = ~key + (key << 15);
		key ^= (key >> 12);
		key += (key << 2);
		key ^= (key >> 4);
		key *= 2057u;
		key ^= (key >> 16);
		return key % nBuckets;
	}
	
	uint32_t BitHash2(uint32_t key, uint32_t nBuckets) 
	{
		key = (key + 0x7ed55d16u) + (key << 12);
		key = (key ^ 0xc761c23cu) ^ (key >> 19);
		key = (key + 0x165667b1u) + (key << 5);
		key = (key + 0xd3a2646cu) ^ (key << 9);
		key = (key + 0xfd7046c5u) + (key << 3);
		key = (key ^ 0xb55a4f09u) ^ (key >> 16);
		return key % nBuckets;
}
\end{lstlisting}

Performance is also critical. Bitwise mixing functions (e.g., Jenkins-style) improve entropy while maintaining low computational cost. Building on this, we designed two GPU-oriented variants, \emph{BitHash1} and \emph{BitHash2}. Both apply sequences of shifts, XORs, and additions to achieve avalanche behavior, then reduce modulo to fit within $|B|$ buckets. 

We also evaluated widely used non-cryptographic functions. MurmurHash~\cite{appleby2008murmurhash} is popular for its strong distribution and high throughput. CityHash \cite{google_cityhash} optimizes for modern architectures, achieving low collision rates with reduced instruction count. CRC-based hashes (CRC-32/CRC-64) \cite{castagnoli1993optimization}, though originally designed for error detection, are attractive on GPUs due to table-based implementations that replace arithmetic with cache-friendly lookups. To analytically characterize how uniformly these hash functions distribute keys, we formalize the expected bucket occupancy and collision behavior under ideal uniform hashing as follows.

	\begin{theorem}[Uniform hashing occupancy and collisions] \label{theorem:uniformhashing}
		Let $n$ keys be mapped independently and uniformly at random into $m$ buckets. For bucket loads $L_b$ and total collisions $Y=\sum_{b=1}^{m}(L_b-1)_+$:
		\begin{align*}
			\Pr[L_b=k] &= \binom{n}{k}\left(\tfrac{1}{m}\right)^k\!\!\left(1-\tfrac{1}{m}\right)^{n-k},\\
			\mathbb{E}[Y] &= n - m\Big(1-\big(1-\tfrac{1}{m}\big)^n\Big),\\
			\Pr(\text{collision for key}) &= 1 - \left(1-\tfrac{1}{m}\right)^{n-1}.
		\end{align*}
		For $n << m$ with $\lambda=n/m$, $L_b \sim \mathrm{Poisson}(\lambda)$, $\mathbb{E}[\text{empty}] \approx m e^{-\lambda}$, and $\mathbb{E}[Y]\approx n^2/(2m)$.
	\end{theorem}

To quantify deviations from uniform hashing, we define the \emph{Collision Speedup Ratio} (CSR):
\[
CSR = \frac{\mathbb{E}[Y]}{Y_{\text{observed}}}.
\]
$CSR$ value of $1$ indicates perfect agreement with uniform hashing, $CSR>1$ indicates fewer collisions than expected (better spread), and $CSR<1$ indicates excess collisions.

\begin{figure}[!ht]
	\centering
	\scalebox{0.19}{\includegraphics{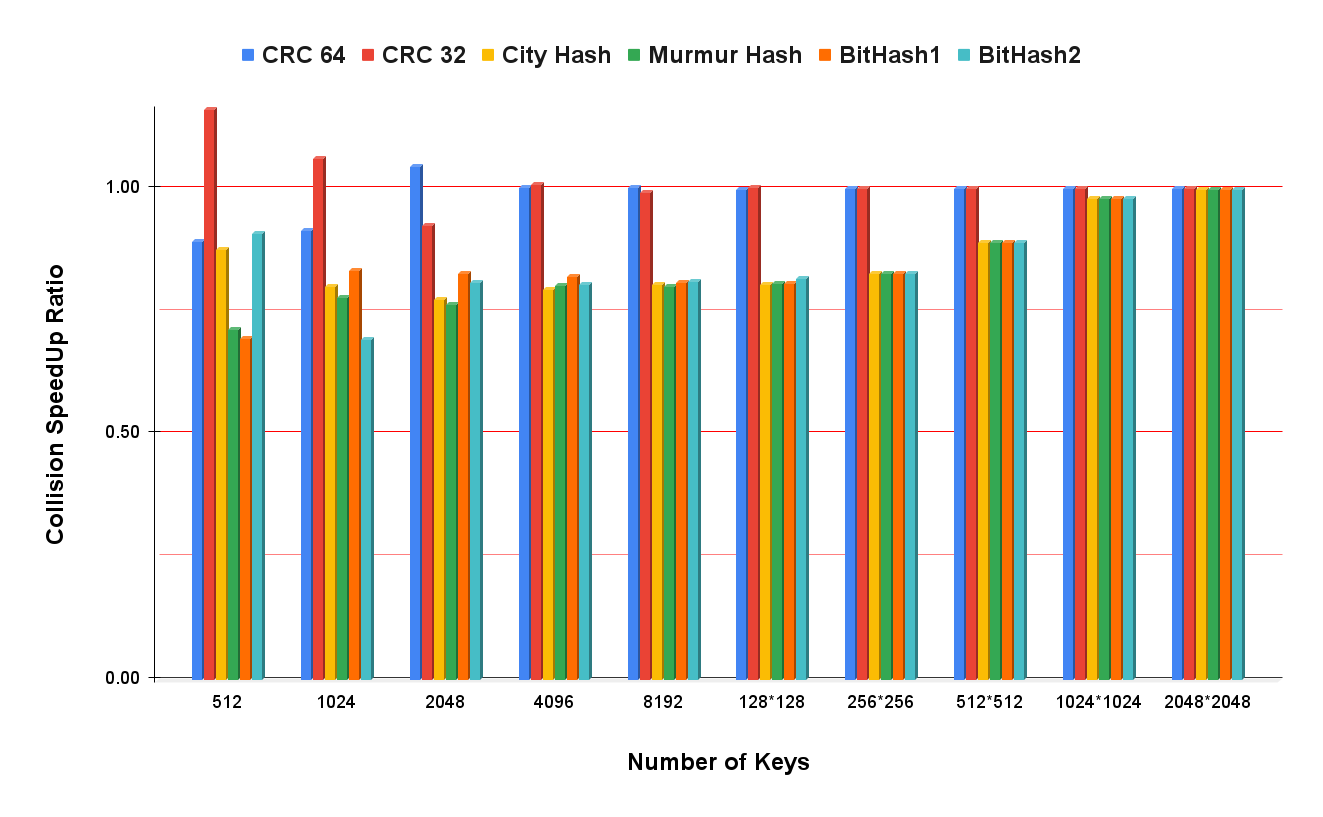}}
	\caption{Collision Speedup Ratio (CSR) of hash functions across varying key counts for $m=512^2$ buckets.}
	\label{fig:HashFunctionCSR}
\end{figure}

We evaluated CRC-32, CRC-64, CityHash, MurmurHash, BitHash1, and BitHash2 hash functions on NVIDIA GeForce RTX 4090 GPU using $m=512^2$ buckets and $n$ ranging from 512 to $2048^2$. Results in Fig.~\ref{fig:HashFunctionCSR} show that at low load factors, deterministic hashes such as BitHash and CityHash exhibit mild clustering ($CSR<1$). As $n$ grows and randomness dominates, their distributions converge to uniform. CRC functions consistently achieve $CSR \approx 1$ across all scales, confirming excellent spread, albeit at higher computational cost due to table lookups. Overall, CRC offers the strongest uniformity, while BitHash variants provide a lightweight trade-off between entropy and throughput, making them also attractive for GPU cuckoo hashing where thousands of hashes must be computed per batch.

\subsection{Supported Concurrent Operations} \label{subsec:HashFunctionsOperations}
\hivehash processes workloads ($S$) as a batch of single operations or concurrent mix of operations. Each key is paired with a single value item. The following operations are supported:

\begin{itemize}
	\item \texttt{Insert($\langle k,v\rangle$)}: Add a new key-value pair to the hash table. $S \leftarrow S \cup {\langle k,v\rangle}$
	\item \texttt{Replace($\langle k,v\rangle$)}: If a key \var{k} already exists, replace its value with an associated value \var{v}. $S \leftarrow (S - {\langle k,*\rangle}) \cup {\langle k,v\rangle}$
	\item \texttt{Delete(k)}: Remove the specific key along with its value from the hash table. $S \leftarrow S - {\langle k,v\rangle \in S}$
	\item \texttt{Search(k)}: Retrieve the value of given key or return $\perp$ if not found.
\end{itemize}

\subsection{Warp-Aggregated-Bitmask-Claim (WABC)} \label{subsec:WABC}
\begin{figure}[!ht]
	\centering
	\scalebox{0.45}
	{\includegraphics{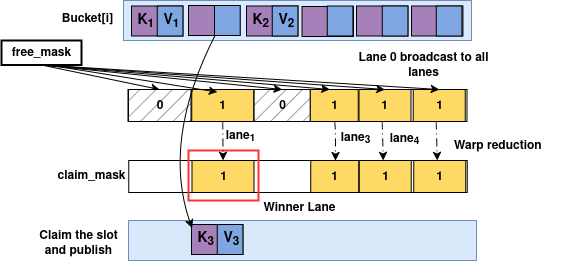}}
	\caption{WABC Approach from Warp's perspective.}
	\label{fig:WABC}
\end{figure}

\hivehash employs a warp-cooperative synchronization protocol to claim slots in constant time, ensuring global consistency by a single, strong atomic read-modify-write (RMW) operation. The approach, illustrated in \fig{WABC}, avoids scanning all 32 KV slots in a bucket, thereby reducing memory traffic and control divergence: only one 32-bit mask is read instead of 32 individual 64-bit slots. Lane~0 first performs a relaxed load of the 32-bit \emph{free\_mask[bucket\_index]}, which typically fits within a single L1/L2 cacheline, and broadcasts the value to all lanes using \texttt{\_\_shfl\_sync}. Each lane~$i$ then checks its corresponding bit to determine whether its slot is free, producing a predicate that indicates slot availability. A warp-wide ballot aggregates those predicates into a \emph{claim mask} identifying all candidate slots. If the mask is non-zero, the lowest-index free lane is elected as the winner, and only that lane performs the atomic RMW to claim the slot and immediately publish the packed KV entry after successful claim.

\subsection{Warp-Cooperative Match-and-Elect (WCME)} \label{subsec:WCME}
Lookup, replace, and delete operations follow a similar warp-synchronous pattern, but instead of consulting the \texttt{freeMask}, each lane coalesced-loads one 64-bit KV entry from the bucket into a thread-local register (\texttt{cached\_kv}), ensuring each slot is fetched exactly once. For a 32-slot bucket, this access pattern maps to two aligned 128-byte memory transactions, fully exploiting GPU memory coalescing. Each lane extracts the key from its KV pair and compares it against the query key. A warp-wide ballot produces a \texttt{match\_mask} identifying the lanes with a match. If the mask is zero, the warp exits early. Otherwise, the first set bit in the mask is selected, and the corresponding lane is elected as the winner. The winner then returns the value (lookup), atomically updates the slot with a new KV (replace), or clears the slot with an empty KV (delete). By combining coalesced loads, register caching, and warp-synchronous election, WCME eliminates redundant traffic and ensures that only one lane performs the critical update while all threads remain active on the same control path.

\section{Implementation Details} \label{sec:Implementation Details}
In this section, we present the implementations of the supported operations of \hivehash on modern NVIDIA GPU hardware. 

\subsection{Insert \& Replace} \label{subsec:InsertAndReplace}
Insertion and replacement follow a four-step strategy designed for maximum concurrency. Different warps can advance through separate phases concurrently without coarse-grained bucket locks in the first two steps. This design separates lock-free fast paths from bounded synchronization phases, sustaining high throughput under contention.

For a given key $k$, the operation first computes hash indices $\{h_1(k), h_2(k), \dots, h_d(k)\}$ where $d$ is the number of configured hash functions (\S\ref{subsec:hash-functions-performance}). Each warp progresses through the following steps, terminating early upon success to reduce table contention. 
\begin{itemize}
	\item \textbf{Step 1:} Checks whether the key already exists in one of its candidate buckets; if found, atomically replace the associated value.
	\item \textbf{Step 2:} If the key is new, claim a free slot using the Warp-Aggregated-Bitmask-Claim (WABC) protocol (\S\ref{subsec:WABC}) in a warp-cooperative fashion.
	\item \textbf{Step 3:} If all candidate buckets are full, select a victim entry and displace it into its alternate bucket via bounded cuckoo eviction.
	\item \textbf{Step 4:} If the eviction bound is exceeded, redirect the entry into the overflow stash to guarantee progress.
\end{itemize}

\subsubsection*{Step 1: Replace} \label{subsubsection:replacePath}
The replace step is triggered when the target key already resides in one of its candidate buckets. It follows the warp-cooperative match-and-elect (WCME, \S\ref{subsec:WCME}) protocol to identify the matching slot and update it atomically. As shown in Algorithm~\ref{alg:tryreplace}, each lane $\ell$ first loads its assigned 64-bit KV entry from bucket $b$ into a thread's private register via a coalesced access (\texttt{cached\_kv}, line~\ref{alg:tryreplace:cachedKV}). Each lane unpacks the key and compares it against the query key $k$ (line~\ref{alg:tryreplace:query_and_compare}). The results are aggregated into a warp-wide match mask using \texttt{\_\_ballot\_sync} \footnote{\texttt{\_\_ballot\_sync}, \texttt{\_\_ffs}, and \texttt{\_\_shfl\_sync} are CUDA warp-level intrinsics~\cite{nvidia_warp_primitives}.}(line~\ref{alg:tryreplace:matchingmask}). If the mask is empty, the warp exits immediately (line~\ref{alg:tryreplace:early_exit}), guaranteeing constant-time failure on key misses. Otherwise, the first set bit in the mask is selected using \texttt{\_\_ffs}, and the corresponding lane is elected as the winner (line~\ref{alg:tryreplace:winner}). Only the winner attempts the update: it constructs a new packed KV and invokes \texttt{compare\_exchange\_strong} on the slot, using the cached value as the expected argument to detect concurrent modifications (line~\ref{alg:tryreplace:cas_update}). Finally, the success flag is disseminated to all lanes via \texttt{\_\_shfl\_sync} (line~\ref{alg:tryreplace:success_broadcast}), ensuring a consistent outcome across the warp. By restricting the update to a single elected lane and issuing at most one CAS per matching key, this step avoids redundant traffic and maintains correctness under concurrency.
\begin{algorithm}[t]
	\caption{\textsc{ReplacePath}$(T,b,k,v,S)$}
	\label{alg:tryreplace}
	\DontPrintSemicolon
	\KwIn{Hash Table $T$, Bucket index $b$, key $k$, value $v$, Number of slots in bucket $S$}
	\KwOut{\textbf{true} on success}
	\BlankLine
	$cached\_kv_\ell \gets T.\textit{buckets}[b][\ell]$ \; \label{alg:tryreplace:cachedKV}
	\textit{// all lanes $\ell$ in warp, $\ell \in [1,S]$}\;
	$match_\ell \gets (\textsc{UnpackKey}(cached\_kv_\ell) == k)$\; \label{alg:tryreplace:query_and_compare}
	$M \gets$ aggregation of $\{match_\ell\}$ across the warp\; \label{alg:tryreplace:matchingmask}
	\If{$M == 0$}{\Return \textbf{false}} \label{alg:tryreplace:early_exit}
	\textit{// elect first matching lane}\;
	$w \gets \textsc{FirstSet}(M)$ \label{alg:tryreplace:winner}
	\BlankLine
	\textit{// winner lane updates the slot}\;
	\If{lane $w$}
	{
		$old \gets cached\_kv$\;
		$new \gets \textsc{Pack}(k,v)$\;
		$success \gets \textsc{CAS}\big(T.\textit{buckets}[b][w],\ old,\ new\big)$\; \label{alg:tryreplace:cas_update}
	}
	\Return broadcast$(success,$ from lane $w)$\; \label{alg:tryreplace:success_broadcast}
\end{algorithm}

\begin{algorithm}[t]
	\caption{\textsc{ClaimThenCommit}$(T,b,kv,W,S)$}
	\label{alg:claimthencommit}
	\DontPrintSemicolon
	\KwIn{Hash table $T$, bucket index $b$, key--value word $kv$, warp $W$, Number of slots in bucket $S$}
	\KwOut{Slot index on success, $-1$ on failure}
	
	$mask \gets \textsc{Load}(T.\textit{freeMask}[b])$ by lane 0, broadcast to $W$\; \label{alg:claimthencommit:loadfreemask}
	$mask \gets mask \,\&\, FULL\_MASK$ \tcp*[r]{mask out unused slots}
	\If{$mask == 0$}{\Return $-1$} \label{alg:claimthencommit:earlyexit}
	
	\BlankLine
	$C \gets$ lanes in $W$ where $(mask \,\&\, (1<< lane)) \neq 0$\; \tcc*{$lane \in [1,S]$} \label{alg:claimthencommit:claim_mask}
	$winner \gets \textsc{FirstSet}(C)$ \tcp*[r]{winner election}
	$claimed \gets -1$\;
	
	\BlankLine
	\If{lane $winner$}{
		$slotBit \gets (1<< winner)$\;
		$old \gets \textsc{FetchAnd}(T.\textit{freeMask}[b], \neg slotBit)$\; \label{alg:claimthencommit:clear_mask}
		\If{$old \,\&\, slotBit \neq 0$}{
			$T.\textit{buckets}[b][winner] \gets kv$ \tcp*[r]{publish new entry} \label{alg:claimthencommit:publish_kv}
			$claimed \gets winner$\;
		}
		\Else{
			\textsc{FetchOr}$(T.\textit{freeMask}[b], slotBit)$ \tcp*[r]{restore bit if failed} \label{alg:claimthencommit:restore_mask}
		}
	}

	\BlankLine
	\Return broadcast$(claimed,$ from lane $winner)$\; \label{alg:claimthencommit:broadcast}
\end{algorithm}
\subsubsection*{Step 2: Claim-Then-Commit} \label{subsubsection:claimPath}
The claim-then-commit step is responsible for inserting a new key–value pair when a free slot is available in one of the candidate buckets. It follows the WABC protocol (\S\ref{subsec:WABC}), but extends it with an immediate commit of the entry. As shown in Algorithm~\ref{alg:claimthencommit}, the first lane (lane~0) performs a relaxed load of the 32-bit \texttt{freeMask} for bucket $b$ and broadcasts (\texttt{\_\_shfl\_sync}) it to all lanes (line \ref{alg:claimthencommit:loadfreemask}). The mask is then intersected with \texttt{FULL\_MASK} to ignore inactive bits. If the resulting mask is empty (line \ref{alg:claimthencommit:earlyexit}), the warp terminates early, indicating the bucket is full. Otherwise, each lane determines whether its slot ($1<< lane$) is available and participates in a warp-wide ballot to aggregate the candidate lanes, eligible to claim (line \ref{alg:claimthencommit:claim_mask}). The first set bit in the ballot result is chosen as the winner. Only the winner lane attempts the claim: it issues an atomic \texttt{fetch\_and} on the mask to clear its bit (line \ref{alg:claimthencommit:clear_mask}). If the old mask confirms ownership of the slot, the lane immediately commits the packed key–value word into the bucket array (line \ref{alg:claimthencommit:publish_kv}). If the claim fails, the bit is restored via \texttt{fetch\_or} (line \ref{alg:claimthencommit:restore_mask}), ensuring correctness. Finally, the index of the claimed slot (or $-1$ on failure) is broadcast back to the warp (line \ref{alg:claimthencommit:broadcast}), so all lanes observe a consistent result. This step provides a deterministic, lock-free, constant-time insertion path, requiring at most one atomic RMW on \texttt{freeMask} and one store to publish the entry.

\subsubsection*{Step 3: Cuckoo Eviction} \label{subsubsection:cuckooEvictionPath}
If Step~2 fails to claim a free slot in either candidate bucket, the warp enters the eviction path (Algorithm~\ref{alg:cuckoo-evict-insert}). 
Each round first re-attempts the lock-free claim; if it still fails, the first lane (lane~0) briefly \emph{acquires} the bucket lock (line \ref{alg:cuckoo-evict-insert:lock-bucket}), performs a relaxed read of \texttt{freeMask} (line \ref{alg:cuckoo-evict-insert:load-free-mask}), and takes one of two actions: (i) if a free bit exists, it claims that bit and \emph{publishes} the newcomer into the corresponding slot, recording the outcome as \textsc{PlacedWithoutEvict} (line \ref{alg:cuckoo-evict-insert:place-without-eviction-start}-\ref{alg:cuckoo-evict-insert:place-without-eviction-end}); (ii) otherwise, it selects the first occupied slot, \emph{observes} the victim entry, overwrites the slot with the newcomer, and records as \textsc{Evicted}$(s)$ (line \ref{alg:cuckoo-evict-insert:eviction-start}-\ref{alg:cuckoo-evict-insert:eviction-end}). Both outcome and (if any) victim are broadcast to the warp (line \ref{alg:cuckoo-evict-insert:broadcast}). On \textsc{PlacedWithoutEvict} the operation terminates successfully; on \textsc{Evicted}$(s)$ the warp re-routes the evicted key to its alternate bucket (line \ref{alg:cuckoo-evict-insert:reroute-start}-\ref{alg:cuckoo-evict-insert:reroute-end}) and repeats, for up to $T.\textit{max\_evictions}$ rounds. This step confines synchronization to a short critical section per round (only for lane~0), keeps read-side traffic relaxed under the lock, and exposes a clear linearization point at the store that publishes the newcomer (either into a free slot or the victim slot).

\begin{algorithm}[!t]
	\caption{CuckooEvictAndInsert$(T,b_0,kv_0,W)$}
	\label{alg:cuckoo-evict-insert}
	\DontPrintSemicolon
	\KwIn{Start bucket $b_0$, packed item $kv_0$, warp $W$}
	\KwOut{\textbf{true} on success}
	$kv \gets kv_0$; \quad $b \gets b_0$; \quad $VALID \gets \textsc{ValidMask}()$\;
	\For{$kick \gets 1$ \KwTo $T.\textit{max\_evictions}$}
	{
		\If{\textsc{ClaimThenCommit}$(T,b,kv,W)$}{\Return \textbf{true}}  \tcp*[r]{lock-free fast path}
		
		$\mathit{outcome} \gets \bot$;\quad $victim \gets \bot$\;
		
		\If{lane $0$}
		{
			\textsc{Lock}$(T,b)$ \tcp*[r]{CAS with \emph{acquire}} \label{alg:cuckoo-evict-insert:lock-bucket}
			\textit{//relaxed load}\;
			$*fm \gets \&(T.\textit{freeMask}[b]) \,\&\, VALID$\; \label{alg:cuckoo-evict-insert:load-free-mask}
			\If{$fm \neq 0$}{ 
				$s \gets \textsc{FirstSet}(fm)$;\quad $bit \gets (1 << s)$\; \label{alg:cuckoo-evict-insert:place-without-eviction-start}
				$new \gets fm \,\&\, \neg bit$\;
				$T.\textit{freeMask}[b] \gets new$\;
				$T.\textit{buckets}[b][s] \gets kv$ \tcp*[r]{publish}
				\textsc{Unlock}$(T,b)$ \tcp*[r]{release}
				$\mathit{outcome} \gets \PlacedWithoutEvict$ \label{alg:cuckoo-evict-insert:place-without-eviction-end}
			}
			\Else{
				$occ \gets \neg fm$\; \label{alg:cuckoo-evict-insert:eviction-start}
				$s \gets \textsc{FirstSet}(occ)$\;
				$victim \gets T.\textit{buckets}[b][s]$\;
				\textit{//swap with newcomer and release slot}\;
				$T.\textit{buckets}[b][s] \gets kv$\;
				\textsc{Unlock}$(T,b)$\; 
				$\mathit{outcome} \gets \textsc{Evicted}(s)$\; \label{alg:cuckoo-evict-insert:eviction-end}
			}
		}
	  $\mathit{outcome} \gets$ broadcast$(\mathit{outcome},$ from lane $0)$;\quad 
 		$victim \gets$ broadcast$(victim,$ from lane $0)$\; \label{alg:cuckoo-evict-insert:broadcast}
 		\BlankLine
 		\If{$\mathit{outcome} = \textsc{PlacedWithoutEvict}$}{\Return \textbf{true}}
 		\textit{// re-route the evicted key to its alternate bucket and continue}\;
 		\If{$\mathit{outcome}$ is $\textsc{Evicted}(s)$}   
 		{
 				$kv \gets victim$;\quad $k' \gets \textsc{UnpackKey}(kv)$\; \label{alg:cuckoo-evict-insert:reroute-start}
 				$b \gets \textsc{AltBucket}(k', b, T.\textit{num\_buckets})$\;
 				\Continue \label{alg:cuckoo-evict-insert:reroute-end}
 		}        
	}
	\Return \textbf{false}
\end{algorithm}

\subsubsection*{Step 4: Overflow Stash}\label{subsubsection:stashPath}
When eviction fails to find space within the bounded displacement limit, entries are redirected to a bounded overflow stash implemented as a lock-free ring buffer in global memory. This final fallback mechanism ensures progress while maintaining bounded memory usage.

The stash protocol operates as follows: Producers first read \texttt{head} (relaxed) and \texttt{tail} (acquire) to check available capacity. If space exists, they atomically reserve a slot using \texttt{fetch\_add} on \texttt{tail} (acq\_rel), compute the ring index as \texttt{tail} $\mod$ \texttt{capacity}, and write the key-value pair to \texttt{keys[index]} and \texttt{values[index]}. If the stash is full, the operation is flagged as pending for deferred reinsertion during the next resize epoch, preserving lock-free progress while maintaining bounded memory overhead.

The stash size is configured as a small fraction (typically 1-2\%) of the main table capacity, providing sufficient buffering for temporary overflow while avoiding unbounded memory growth. During table expansion, stashed entries are reprocessed and reinserted into the enlarged table, ensuring eventual consistency and maintaining the hash table's correctness guarantees.
\subsection{Lookup \& Delete}
Both operations leverage the warp-cooperative match-and-elect scheme (WCME, \S\ref{subsec:WCME}) across the $d$ candidate buckets of key $k$ and then perform exactly one winner-only action. For each bucket, \textsc{ScanBucketAndDelete} (Algorithm~\ref{alg:scan-del}) has all $\ell$ lanes coalesced-load and compare their entries (lines~\ref{alg:scan-del:load-and-compare-a}--\ref{alg:scan-del:load-and-compare-b}), form a warp-wide match mask (line~\ref{alg:scan-del:ballot_mask}), and exit immediately on an empty mask (line~\ref{alg:scan-del:early-exit}). Otherwise, the first set bit elects a unique winner (line~\ref{alg:scan-del:winner}). From that point, \emph{lookup} simply returns the value from the winner’s slot, while \emph{delete} performs one CAS to write \textsc{EMPTY} (line~\ref{alg:scan-del:cas}); on success it publishes the vacancy by setting bit $w$ in \texttt{freeMask[$b$]} (line~\ref{alg:scan-del:publish-mask}), and the result is broadcast so all lanes agree (line~\ref{alg:scan-del:broadcast}). Because $d$ and $S$ are fixed configuration constants, per-key lookup/delete time is \BigO 1 with respect to table size.

\begin{algorithm}[t]
	\caption{\textsc{ScanBucketAndDelete}$(T,b,k,W)$}
	\label{alg:scan-del}
	\DontPrintSemicolon
	\KwIn{Table $T$, bucket index $b$, key $k$, warp $W$, Number of slots in bucket $S$}
	\KwOut{\textbf{true} on success}
	\BlankLine
	\textit{// parallel scan over the bucket}\;
	\textit{//all lanes $\ell$ in warp, $\ell \in [1,S]$ }\;
	{
		$kv_\ell \gets T.\textit{buckets}[b][\ell]$ \; \label{alg:scan-del:load-and-compare-a}
		$match_\ell \gets (\textsc{UnpackKey}(kv_\ell) == k)$\; \label{alg:scan-del:load-and-compare-b}
	}
	$M \gets$ ballot of $\{match_\ell\}$ across $W$\; \label{alg:scan-del:ballot_mask}
	\If{$M == 0$}{\Return \textbf{false}} \label{alg:scan-del:early-exit}
	$w \gets \textsc{FirstSet}(M)$ \tcp*[r]{elect first matching lane} \label{alg:scan-del:winner}
	\BlankLine
	\textit{// winner clears the slot with a single CAS, then frees the bit}\;
	\If{lane $w$}
	{
		$old \gets kv_w$;\;
		$success \gets \textsc{CAS}\!\left(T.\textit{buckets}[b][w],\, old,\,  \textsc{EMPTY}\right)$\; \label{alg:scan-del:cas}
		\If{$success$}{
			set bit $w$ in $T.\textit{freeMask}[b]$ \tcp*[r]{publish free slot} \label{alg:scan-del:publish-mask}
		}
	}
	\Return broadcast$(success,$ from lane $w)$\; \label{alg:scan-del:broadcast}
\end{algorithm}

\subsection{Dynamic Resizing Strategy}\label{subsec:dynamic-resize}

\hivehash enables dynamic capacity scaling without global rehashing by adopting a warp-parallel variant of \emph{Linear Hashing}. 
The table dynamically expands or contracts in a fixed $K$-bucket batch according to its load factor. 
When load factor exceeds $0.9$, it triggers \emph{expansion (\S\ref{subsubsec:expansion})}; when it falls below $0.25$, it triggers \emph{contraction} (\S\ref{subsubsec:contraction}). 
This adaptive policy maintains balanced occupancy and high throughput while amortizing resizing overhead.

Dynamic resizing introduces two new control fields into \hivehash data structure: 
(i) an \textbf{index mask} $(2^m-1)$ representing the current hashing round, and 
(ii) a \textbf{split pointer} tracking how many low buckets have been split. 
These enable independent per-bucket growth and shrinkage across multiple rounds, avoiding full-table rebuilds.

\subsubsection{Expansion (Split Phase)}\label{subsubsec:expansion}

During expansion, \hivehash allocates $K$ new buckets and splits $K$ existing ones, starting at
$\texttt{split\_ptr}, \texttt{split\_ptr+1}, \dots, \texttt{split\_ptr+K-1}$.
Each source bucket at index $b_{\text{src}}$ is paired with a new
partner bucket $b_{\text{dst}} = b_{\text{src}} + 2^m$, where $m$ is the current round level.
Each warp cooperatively processes one $(b_{\text{src}}, b_{\text{dst}})$ pair.

Each lane reads coalesced 64-bit slot from the source bucket and determines whether it should
\emph{stay} or \emph{move} to its partner bucket (\emph{b\_dst}).
The decision uses one extra hash bit from the next round mask based on whether the next hash bit directs the key to new buckets:
\begin{lstlisting}[style=codeC]
	next_mask = (table->index_mask << 1) | 1u;
	should_move = ((hash(key) & next_mask) == b_dst);
\end{lstlisting}
Entries satisfying this condition are considered movers.
The warp then forms a compact group of movers using a ballot and prefix sum:
\begin{lstlisting}[style=codeC]
	move_mask = W.ballot(should_move);
	my_rank   = __popc(move_mask & 
	((1u << lane) - 1));
	if (should_move) {
		dst->kv[my_rank] = kv;   // compacted placement
		src->kv[lane]    = EMPTY;
	}
\end{lstlisting}
Lane~0 finally updates both free masks, marking released slots in the
source bucket and occupied slots in the partner bucket:
\begin{lstlisting}[style=codeC, numbers=left]
	src_mask |= move_mask;                
	dst_mask &= ~((1u << n_movers) - 1);
\end{lstlisting}
When all lower buckets ($2^m$) are split, the table advances to the next
round by doubling its addressable range:
\begin{lstlisting}[style=codeC, numbers=left]
	index_mask = (index\_mask << 1) | 1;
	split_ptr = 0;
\end{lstlisting}

Only buckets currently being split participate, yielding \BigO{K}  migration
cost per expansion step.

\subsubsection{Contraction (Merge Phase)} \label{subsubsec:contraction}
When the table’s load factor drops below the lower threshold, \hivehash contracts symmetrically by merging $K$ partner buckets back into their base buckets. Each warp processes one pair $(b_{\text{dst}}, b_{\text{src}} = b_{\text{dst}} + 2^m)$. Each lane scans one slot of the source bucket, and movers are determined by a warp-wide ballot:
\begin{lstlisting}[style=codeC, numbers=left]
	kv = src->kv[lane];
	live   = (kv != EMPTY);
	occ_mask = W.ballot(live);
	my_rank = __popc(occ_mask & ((1u << lane) - 1));
\end{lstlisting}
The destination bucket’s free slots are enumerated, and each mover selects the $r$-th free slot using a prefix-rank mapping:
\begin{lstlisting}[style=codeC, numbers=left]
	pos = select_nth_one(dst_freeMask, my_rank);
	dst->kv[pos] = kv; src->kv[lane] = EMPTY;
\end{lstlisting}
If the destination bucket lacks sufficient free slots ($n_\text{move} > n_\text{free}$), the merge aborts early. Otherwise, lane~0 updates the free masks atomically:
\begin{lstlisting}[style=codeC, numbers=left]
	src_freeMask = 0xFFFFFFFFu;           
	dst_freeMask &= ~used_mask;    
\end{lstlisting}
After $2^m$ merges, the round regresses:
\[
\texttt{index\_mask} >>= 1,\quad
\texttt{split\_ptr} = \texttt{index\_mask}+1.
\]

\section{Experimental Setup \& Performance Evaluation} \label{sec:Experimental Setup and Perf Eval}

This section outlines the experimental environment and evaluation methodology used to assess the performance of \hivehash. We evaluate \hivehash across diverse scenarios, including balanced and imbalanced workloads, and stress tests at extreme load factors using synthetic datasets of up to 32 million uniformly distributed KV pairs.

The \hivehash kernel follows a \emph{warp-centric} execution model: each warp cooperatively executes one operation (insert, lookup, or delete) in a warp-synchronous fashion. This mapping minimizes divergence, confines synchronization to warp scope, and ensures consistent performance even under irregular workloads. Our PTX\footnote{Parallel Thread Execution (PTX) is an intermediate virtual ISA that serves as the compilation target of CUDA kernels before translation to device-specific machine code.} analysis reports a register usage of 90 with no register spilling, confirming that the kernel achieves high occupancy on Ada-class GPUs. To sustain throughput, \hivehash employs a bucketed two-choice placement policy with warp-wide probing (\S\ref{subsec:hash-functions-performance}) and lightweight metadata for fine-grained concurrency control.

\subsection{Hardware Environment} \label{subsec:Hardware Environment}
All experiments were conducted on an NVIDIA GeForce RTX~4090 GPU featuring 16,384 CUDA cores, 24~GB of GDDR6X memory, and 1,008~GB/s of peak memory bandwidth. The device is based on the Ada Lovelace architecture (compute capability~8.9) and was evaluated using the CUDA~12.2 toolkit.
Kernels were compiled with full optimization (\texttt{-O3}) and tuned for maximum occupancy using the CUDA occupancy calculator \cite{nvidia_occupancy_calculator}. All measurements were averaged over ten runs after a warm-up phase to remove initialization overhead. Performance is reported in millions of operations per second (MOPS) or giga ($10^9$) operations per second (GOPS), measured using CUDA event timers with millisecond precision.

To evaluate the efficiency of the dynamic resizing mechanism (\S\ref{subsec:dynamic-resize}), we also measure the throughput of bucket expansion and contraction. On the RTX~4090, \hivehash achieves 16.8 ~GOPS during expansion and 23.7~GOPS during contraction when processing 32,768 buckets—approximately 3–4× faster than SlabHash under identical conditions - indicating resizing overhead  negligible relative to steady-state hash table.

\subsection{Hash Functions Performance} \label{subsec:hash-functions-performance}
We evaluated the performance of different combinations of hash functions within \hivehash under concurrent insert-only workloads. The objective is to assess the trade-offs between \textit{lookup-based (CRC32 \& CRC64)} and \textit{computation-based (BitHash1, BitHash2, CityHash, MurmurHash)} hash functions to quantify the performance impact of using two versus three hash functions during insertion. The lookup-based functions use precomputed lookup tables stored in GPU constant memory, which is read-only and cached for all threads within a warp. 

\begin{figure}
	\centering
	\resizebox{0.45\textwidth}{!}{\includegraphics[scale=1.0, height=\textheight]{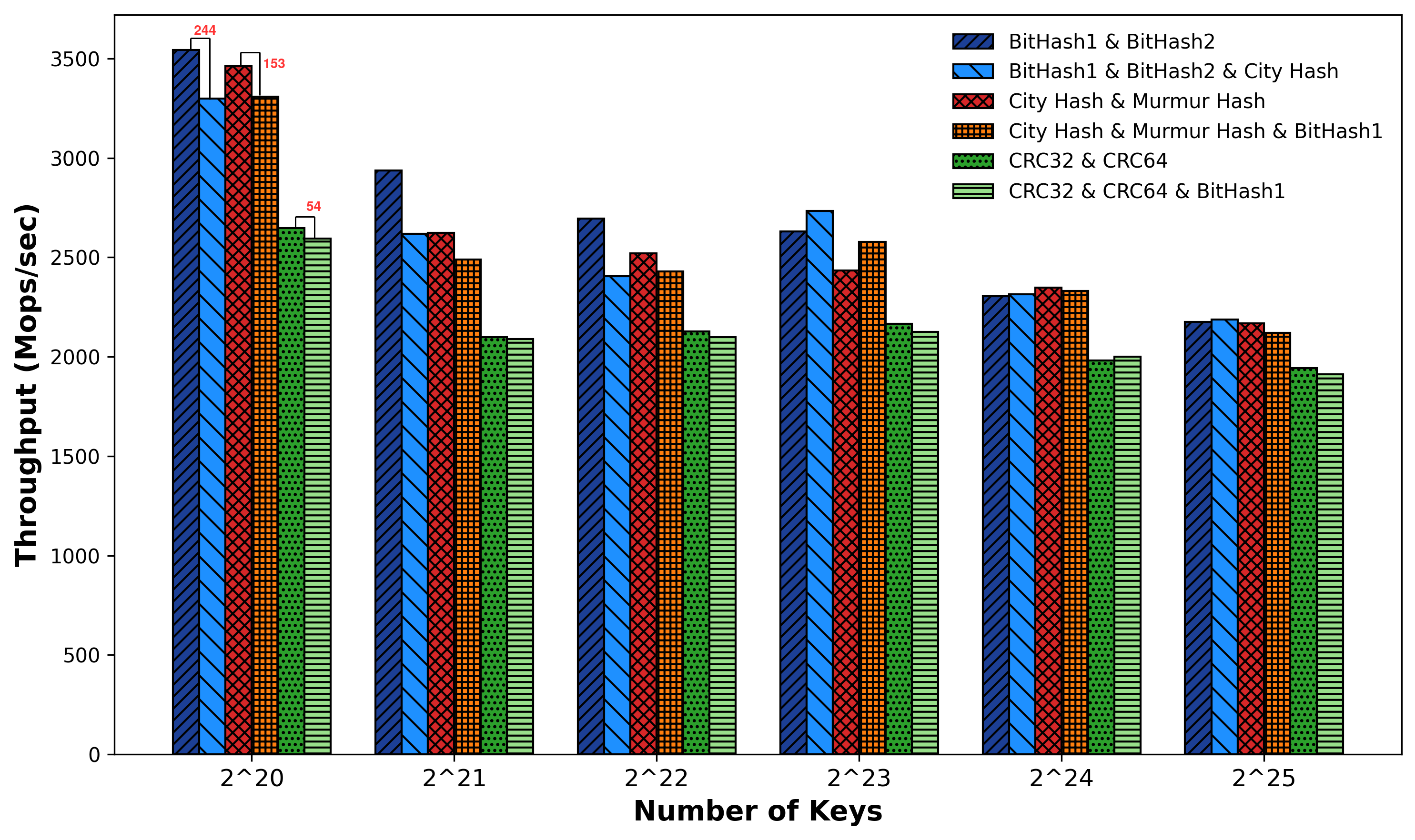}}
	\vspace{-4pt}
	\caption{Insertion throughput comparison of lookup-based and computation-based hash function pairs in \hivehash. Results show two-hash configurations consistently outperform three-hash alternatives.} \label{fig:hash_functions_perf}
\end{figure}

 \fig{hash_functions_perf} presents insertion throughput of six hash function combinations as the number of unique keys varies from $2^{20}$ to $2^{25}$.
Across all input sizes, two-hash configurations consistently outperform their three-hash counterparts, as the marginal improvement in key distribution uniformity does not justify the additional computational cost. The BitHash1 \& BitHash2 pair achieves the highest throughput of 3,543 MOPS, while adding a third hash function (CityHash) reduces performance by 244 MOPS.

Although the theoretical collision speed-up ratio (\fig{HashFunctionCSR}) predicted that lookup-based CRC functions yields near-uniform key distribution, in practice they exhibit a 12-25\% performance drop due to amplify the higher computational overhead under high concurrent accesses. Therefore, we adopt BitHash1 \& BitHash2 as the default configuration for their superior throughput and balanced performance for all subsequent experiments.

\subsection{Comparative Evaluation with State-of-the-Art Hash Tables} \label{subsec:baseline_comparision}
We evaluate \hivehash against three open-source, state-of-the-art GPU hash tables: WarpCore \cite{junger2020warpcore}, SlabHash \cite{ashkiani2018dynamic}, and DyCuckoo \cite{li2021dycuckoo} under identical experimental conditions.

The evaluation focuses on concurrent throughput under two workload categories: (i) \textit{balanced workloads} consisting of homogeneous operations (bulk insert/replace or bulk lookup), and (ii) \textit{imbalanced workloads} with mixed insert/replace, lookup, and delete operations. All tests use uniformly distributed key-value pairs with performance measured at each system's maximum achievable load factor (SlabHash: 0.92, WarpCore: 0.95, DyCuckoo: 0.9, \hivehash: 0.95). Performance metrics represent averages over ten consecutive runs.


\subsubsection{Balanced Workload} \label{subsubsec:balanced_workload}
In balanced workloads, all operations are homogeneous, consisting of either all insert/replace or all lookup operations. We concurrently insert up to $2^{25}$ unique key-value pairs into an initially empty table until reaching maximum achievable load factors. For bulk query workloads, we retrieve the values of unique query keys from a pre-filled hash table. 

\begin{figure}
	\centering
	\resizebox{0.45\textwidth}{!}{\includegraphics[scale=1.0, height=\textheight]{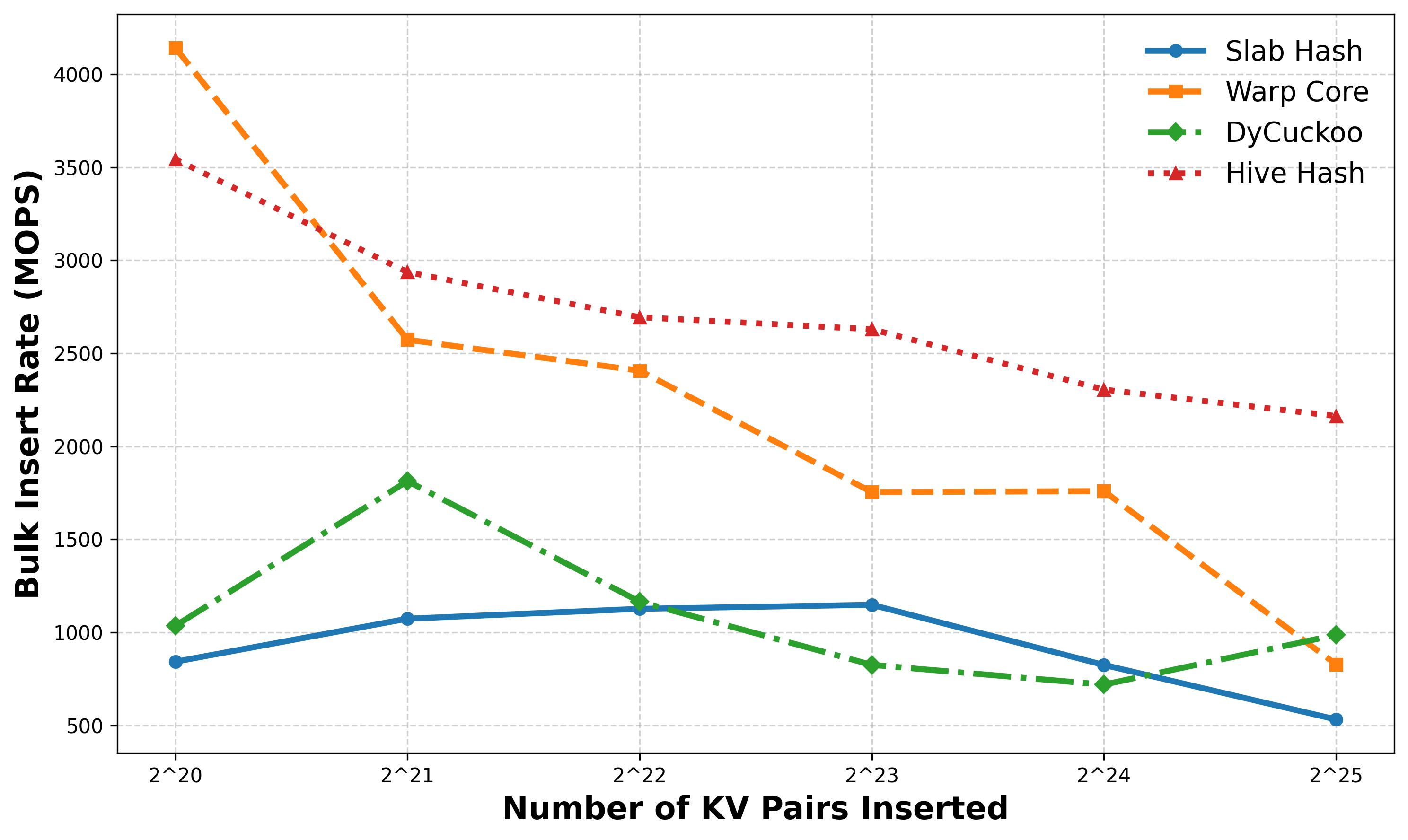}}
	\vspace{-4pt}
	\caption{Performance of concurrent bulk insertion.}
	\label{fig:bulk_insert_perf}
\end{figure}

\fig{bulk_insert_perf} shows the results of bulk insertions. \hivehash consistently delivers the highest throughput (3,543 - 2,162 MOPS) even under high concurrency, outperforming WarpCore and DyCuckoo by $2.5\times$ and SlabHash by $4\times$. Unlike other designs, scaling the number of keys in \hivehash does not increase per-operation control overhead because both WABC and WCME maintain constant execution cost per bucket. As the number of buckets grows, warps access distinct memory regions of the hash table, which naturally reduces inter-warp contention. In contrast, SlabHash introduces linked-list allocator overhead that creates more irregular global memory access patterns with larger key-value pairs. DyCuckoo's uncoordinated multi-round cuckoo relocation increases latency under heavy workloads.

\begin{figure}
	\centering
	\resizebox{0.45\textwidth}{!}{\includegraphics[scale=1.0, height=\textheight]{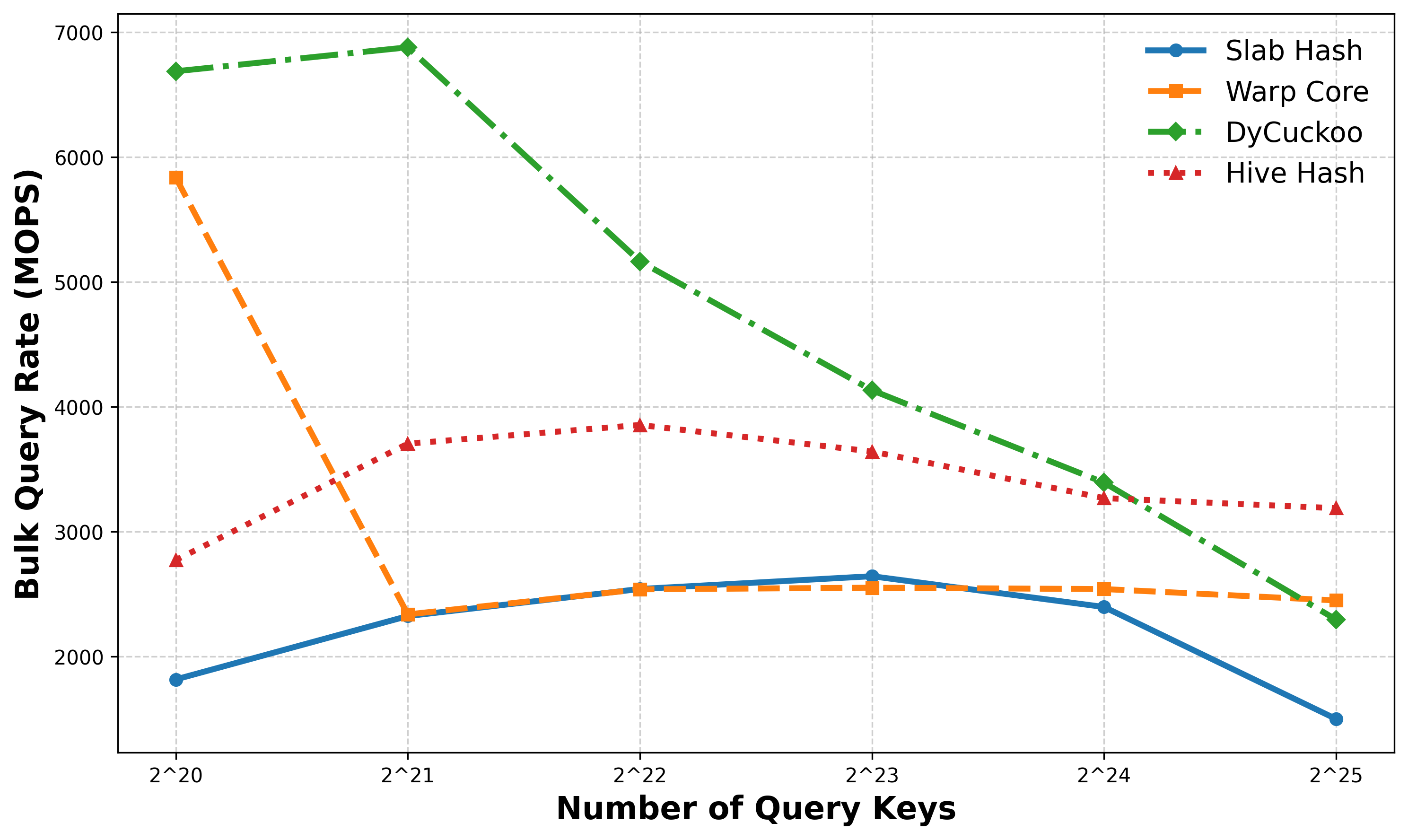}}
	\vspace{-4pt}
	\caption{Performance of concurrent bulk query.}
	\label{fig:bulk_query_perf}
\end{figure}

Figure~\ref{fig:bulk_query_perf} shows the results of bulk queries. \hivehash sustains up to 3,853~MOPS, the highest among all competitors. DyCuckoo performs competitively for smaller tables ($2^{20}$ keys) but its query throughput declines sharply as the table scales, because each lookup must probe all $d$ independent subtables to locate a key. This multi-subtable probing introduces additional global memory accesses and warp divergence, making queries increasingly expensive at large scales. WarpCore and SlabHash exhibit relatively stable but lower throughput: WarpCore is constrained by per-thread atomic synchronization during bucket probing, while SlabHash suffers from pointer-chasing overhead across linked slabs. In contrast, \hivehash leverages a warp-synchronous WCME protocol with a coalesced 64-bit bucket layout, enabling each warp to perform a single aligned probe per candidate bucket. This design minimizes divergence and global memory traffic, sustaining high throughput and predictable latency even under high occupancy.

\subsubsection{Imbalanced Workload} \label{subsubsec:imbalance_workload}
We also benchmarked \hivehash against SlabHash and DyCuckoo using an imbalanced workload with mixtures of insert, lookup, and delete. WarpCore was excluded from this experiment because its per-thread atomic model lacks coordination across warps, often leading to race conditions and ABA problems when insertions and deletions occur concurrently.

\begin{figure}
	\centering
	\resizebox{0.45\textwidth}{!}{\includegraphics[scale=1.0, height=\textheight]{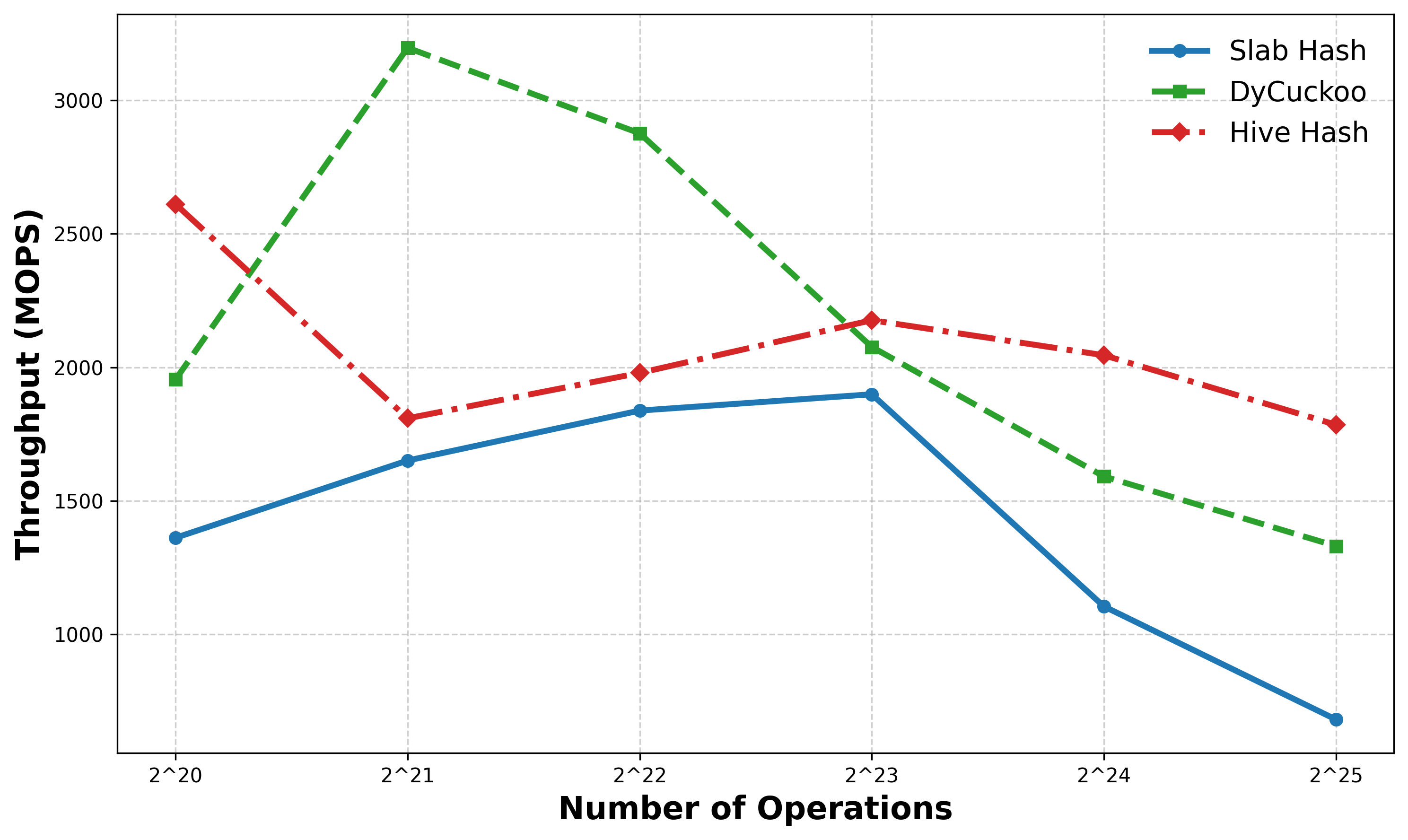}}
	\vspace{-4pt}
	\caption{Performance of imbalanced workload.}
	\label{fig:imbalance_workload_perf}
\end{figure}

\fig{imbalance_workload_perf} shows the result of the insert:lookup:delete ratio 0.5:0.3:0.2. Across all tested cases, \hivehash maintains stable performance after a brief decline at smaller sizes, sustaining throughput from approximately 2,611~MOPS down to 1,796~MOPS as the number of operations increases. This stability arises from Hive hash table's \emph{warp-synchronous update}, which confines synchronization to a single warp per bucket and minimizes inter-warp contention. Each warp operates cooperatively on a bucket, reducing global atomic pressure and eliminating uncoordinated slot ownership. 

SlabHash initially scales modestly but its throughput collapses beyond roughly $2^{23}$ operations due to heavy allocator contention and linked-slab traversal overhead. The pointer-chasing structure of its dynamic slabs leads to non-coalesced memory access and high latency under dense workloads. DyCuckoo performs competitively at smaller scales (peaking near $2^{21}$ keys) due to its shallow probe depth and parallel bucket selection, but performance declines sharply as load increases. The lack of bounded relocation and uneven subtable utilization causes frequent eviction cascades, resulting in significant variance in latency and degraded throughput at large table sizes.

In contrast, \hivehash sustains superior throughput and robustness under mixed workloads by efficiently coordinating concurrent updates and reusing freed slots without global rehashing. 

\subsection{Insertion Step Breakdown}
\label{subsec:insertion_breakdown}

To better understand the behavior of \hivehash under varying contention levels, we analyzed the time contribution of each insertion step with the load factor increasing. This breakdown helps identify performance bottlenecks and guides optimization for high-occupancy scenarios.

For in-kernel timing, we use \texttt{clock64()} \cite{guide2024cuda} to measure fine-grained latency at warp granularity. Each lane records its local start time, and the minimum among all lanes is computed using a warp-level reduction, yielding the earliest start timestamp for the warp. For the end timestamp, all lanes attempt to write to a private register, and the thread writing last represents the maximum completion time within that warp. This technique accurately captures the elapsed clock cycles of each insertion stage while accounting for overlapping warp execution on modern NVIDIA architectures (e.g., RTX 4090 with four concurrent warps per SM). The elapsed time is computed from cycle counts using the GPU's nominal clock frequency of 2.52 GHz.

\begin{figure}[!t]
	\centering
	\resizebox{0.47\textwidth}{!}{
		\includegraphics[scale=1.0]{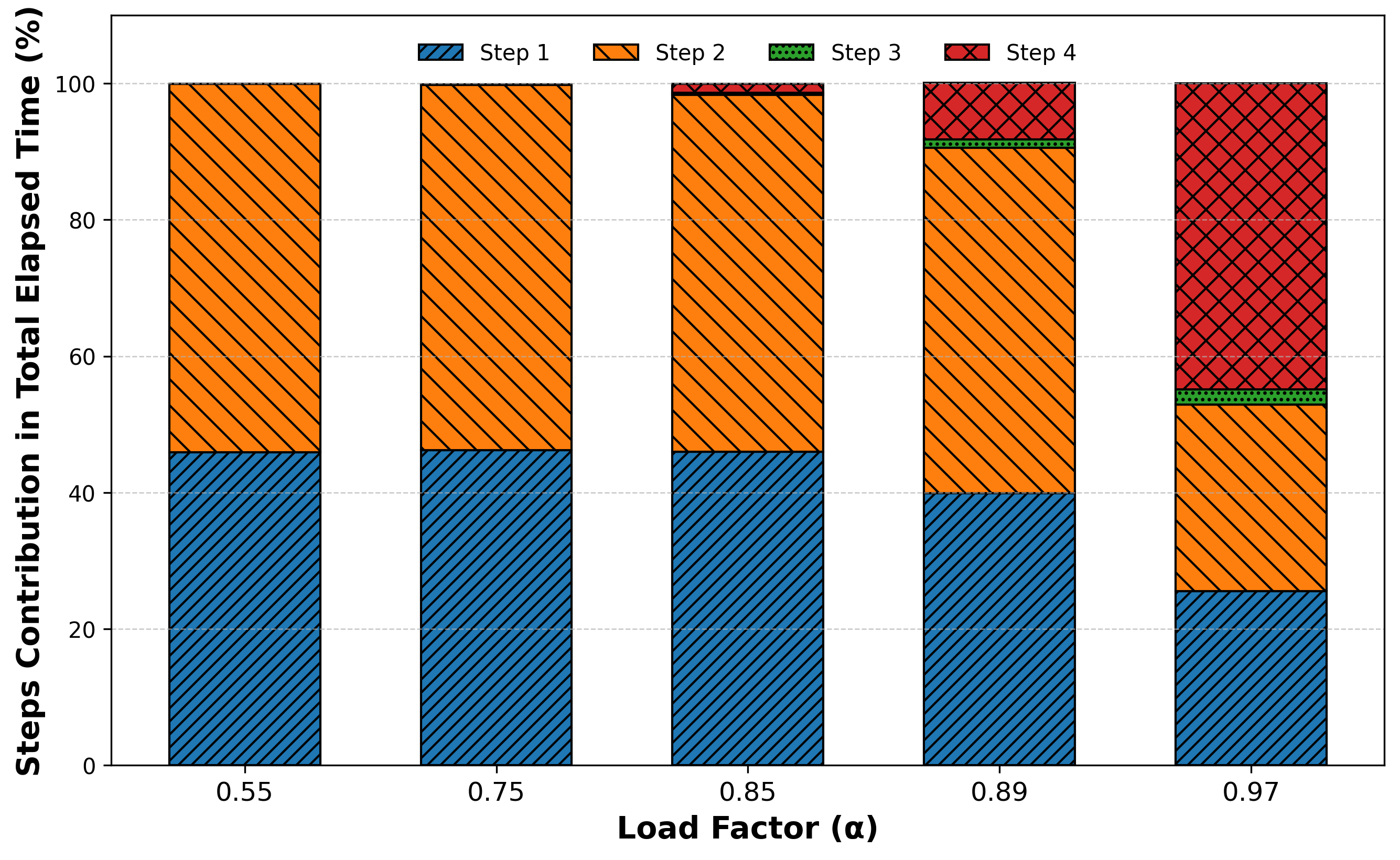}}
	\vspace{-4pt}
	\caption{Insertion step time contribution across varying load factors. Each bar shows the percentage of total elapsed time spent in \textit{Replace} (Step 1), \textit{Claim–then–Commit} (Step 2), \textit{Cuckoo Eviction} (Step 3), and \textit{Stash Fallback} (Step 4).}
	\label{fig:insertion_breakdown}
\end{figure}

\fig{insertion_breakdown} shows the relative time contribution of each insertion step as load factor varies from 0.55 to 0.97. At low occupancy (0.55–0.75), insertions predominantly complete in steps~1 and~2, corresponding to the fast \emph{Replace} and \emph{Claim–then–Commit} steps, which together account for over 95\% of the total elapsed time. Step 3 (\emph{Cuckoo Eviction}) contributes only 0.02–2.2\% of total time, demonstrating that eviction overhead is both infrequent and limited by the fixed eviction bound, thereby confirming the design’s bounded recovery cost under contention. As the table approaches saturation ($\alpha > 0.9$), step~4 (\emph{Stash Fallback}) becomes dominant, consuming nearly half of total time (~41\%) at $\alpha = 0.97$. This surge reflects the growing cost of managing overflow insertions when no vacant slots remain within primary buckets.

Overall, the results highlight that \hivehash maintains low insertion overhead up to 85–90\% occupancy, after which stash handling and collision resolution begin to dominate. These insights motivate initiating table expansion at a load factor threshold of 0.9 to prevent performance degradation near saturation.

\section{Conclusion}  \label{sec:Conclusion}
\hivehash achieves up to 4× higher throughput than existing GPU hash tables across balanced and mixed workloads, sustaining over 95\% load factor without degradation. Its warp-synchronous design ensures stable performance under contention, while dynamic resizing operates 3–4× faster than prior approaches with negligible overhead. Profiling confirms that most insertions complete in lock-free phases and eviction costs remain bounded, validating the efficiency of its concurrency pipeline. The \hivehash library and benchmark suite will be released as open source upon acceptance to support reproducibility and further GPU data-structure research.

\bibliographystyle{IEEEtran}
\bibliography{hive_hash}

\end{document}